\documentclass{mn2e}  
\usepackage{amssymb}
\usepackage{txfonts}
\usepackage{txfonts}
\usepackage[figuresright]{rotating} 
\input epsf.sty  
\newif\ifAMStwofonts  
\AMStwofontstrue

\begin{document}  
  
\title{The pattern of accretion flow onto Sgr A*}  
\author[Mo\' scibrodzka et al.]  
{Monika Mo\' scibrodzka$^1$, Tapas K. Das$^2$, Bozena Czerny$^1$\\  
 $^1$N. Copernicus Astronomical Center, Bartycka 18, 00-716 Warsaw, Poland\\ 
 $^2$ Harish Chandra Research Institute, Allahabad 211 019, India\\
mmosc@camk.edu.pl, tapas@mri.ernet.in, bcz@camk.edu.pl
 } 
\maketitle  
\begin{abstract}  
The material accreting onto Sgr A* most probably comes from the nearby stars.
We analyze the pattern of this flow at distances of a fraction of a parsec 
and we argue that the net angular momentum
of this material is low but non-negligible, and the initially supersonic disk
accretion changes into subsonic flow with constant angular momentum. 
Next we estimate the flow parameters
at a distance $R_{BHL}$ from the black hole and we argue that for the plausible
 parameter
range the accretion flow is non-stationary. The inflow becomes supersonic at
distance of $\sim 10^4 R_g$ but the solution does not continue below the horizon and
the material piles up forming a torus, or a ring, at a distance of
a few up to tens of Schwarzchild radii. Such a torus is known to be unstable
and may explain strong variability of the flow in Sgr A*. Our considerations
show that the temporary formation of such a torus seems to be unavoidable.
 Our best fitting model predicts a rather large accretion
rate of around $4 \cdot 10^{-6}$ $M_{\odot}/ yr$ directly on Sgr A*. We argue that
magnetic fields in the flow are tangled and this allows our model to overcome
the disagreement with the Faraday rotation limits.

\end{abstract}  
  
  
\section{Introduction}
\label{intro}

The center of our Galaxy harbors a massive black hole, and the surrounding
region including the central SMBH is now customary referred as
Sgr A* as a whole, after the radio source first discovered
at that location (for a review, see e.g. Melia \& Falcke 2001).
Sgr A* shows frequent flares originating 
from the direct vicinity of the central
black hole. In the X-ray band, multiple flares are superimposed on
a steady, extended emission at the level of $\sim 2.2 \times 10^{33}$ 
erg s$^{-1}$ cm$^{-2}$. Emission of two extremely bright flares 
have been reported 
so far (Baganoff et al. 2001, Porquet et al. 2003; maximum flux of
$1.0 \pm 0.1 \times 10^{35}$ and $3.6^{+0.3}_{-0.4} \times 10^{35}$,
 respectively), along with 
many fainter flares
observed in the Chandra data (e.g. Eckart et al. 2004, Belanger 
et al. 2005). The duration
of the flares ranges from half an hour to several hours, while 
the rise/decay
time is found to be of the order of few hundred seconds (Baganoff 2003).

Variable emission was also detected in the NIR band. Quiescence
emission takes place at the level of $\sim 1.9$ mJy (Eckart et al. 2004). 
Variable and quiescent emission
was reported by Genzel et al. (2003a) based on the VLT observations, 
and by Ghez et al. (2004) based on Keck data.
In two of the events, a 17 min periodicity was found (Genzel et al. 2003a).
X-ray and NIR outbursts are directly related, as shown by the detection
X-ray and NIR outbursts are directly related, as shown by the detection
of simultaneous NIR/X-ray events (Eckart et al. 2004, 2005). 
The duration of events is of order of tens of minutes. 

The properties of the flare emission suggest that such flares 
originate in the innermost part of accretion flow onto the central 
black hole, and we need to introduce an appropriate theoretical 
model of accretion to understand these flares. The effort is going 
in different directions. Significant effort is devoted to modeling 
the (time-dependent) emissivity of the nuclear region but without specific 
reference
to geometry of the flow (e.g. Liu et al. 2004,2005). On the other
hand, models involving the description of the accretion process 
onto the SMBH of Sgr A$^*$ usually adopt simple geometries like 
purely spherical accretion (Melia 1992, Quataert 2004, Mo\' scibrodzka 2005), 
Bondi-Hoyle accretion (Ruffert, Melia 1994),
high angular momentum ADAF type flows (Abramowicz et al. 1995, Narayan,
Yi \& Mahadevan 1995; Yuan, Quataert \& Narayan 2003, 2004), 
with the most advanced being the coupled disk inflow -- jet-like outflow 
(e.g. Markoff et al. 2001, Yuan, Markoff \& Falcke 2002). Some 
authors addressed the issue of the specific sources of the accreting material:
accretion from the nearest O-type stars during their passage close
to the Sgr A* was discussed by Loeb (2004) and the 3-D hydrodynamical
model of accretion from many close stars was studied by Rockefeller
et al. (2004) and 
Cuadra et al. (2005a,b) but such hydrodynamical 
studies do not have high resolution close to the black hole 
and show the flow at distances above 2'' and 0.1'', correspondingly. 
This last paper
include the motion of the donor stars and stress the importance of the
angular momentum, which is predicted to be small but non-negligible.

In the present paper we estimate the most plausible range of the
angular momentum of the accreting mater by comparing various donor
stars. Since the resulting angular momentum is low we next
apply the low angular momentum flow model which is expected
to produce multi-transonic behavior with standing shocks 
(see e.g. Das 2002, hereafter D02; Das, Pendharkar \& Mitra 2003,
hereafter DPM; Das 2004, Barai, Wiita \& Das
2004, and references therein). We show that for the
estimated parameters of the flow in Sgr A* stationary
solution do not exist and the flow pattern consist of 
a semi-stationary flow above a few Schwarzschild radii, and
an unstable inner ring/torus.

\section{Source of the accreting material}
\label{sect:winds}

\subsection{Stellar winds in the central cluster} 
\begin{figure}  
\epsfxsize = 90 mm  
\epsfbox{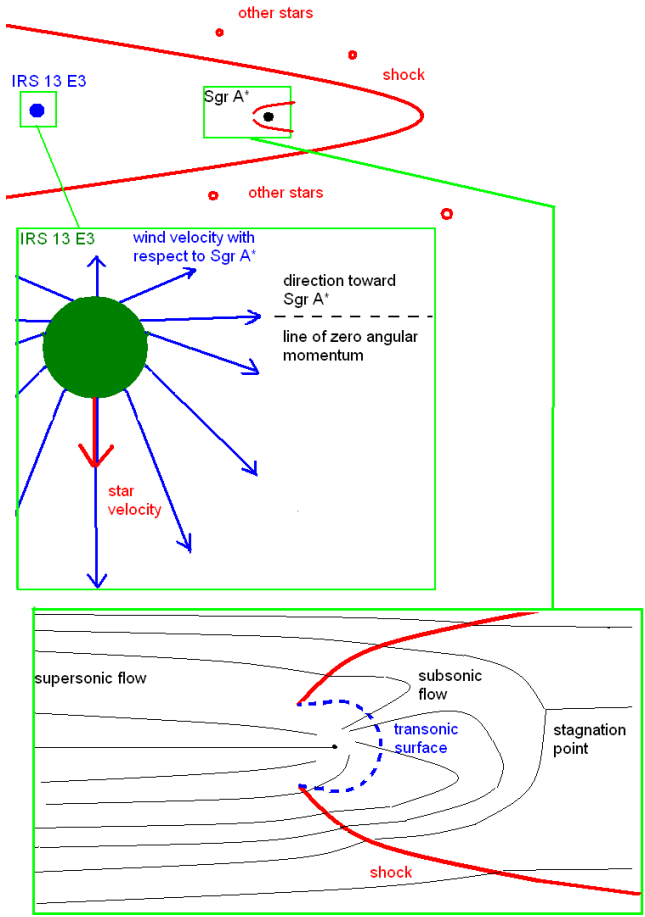} 
\caption{The schematic picture of the wind flow from the star IRS 13E3
towards the Galactic center.  
\label{fig:draw1}}  
\end{figure}
  
Stellar winds originating from the central cluster are plentiful sources of 
gas. Most of this material is likely to be expelled from the central region
(e.g. Quataert 2004) but a remaining fraction may power the observed 
activities.
Among various mass loosing stars, particularly active one is the 
Wolf-Rayet star
IRS 13 E3 (Paumard et al. 2001, Melia \& Falcke 2001), resolved to consist 
actually of two dusty Wolf-Rayet stars (Maillard et al. 2004). 
The potential role of each source in dominating the flow pattern can be 
estimated through its ram pressure (e.g. Melia 1992). 

We consider the stars listed by Rockefeller et al. (2004) as potential important sources 
of the material. For each of the stars independently we calculate the ram pressure,
 $0.5 \rho_{wind} v_{wind}^2$  at the location of Sgr A*, where $\rho_{wind}$ is the
 density of the stellar wind and $v_{wind}$ is the wind velocity. Wind density can 
be found knowing the outflow rate, $\dot M_{wind}$ and the distance, $D_*$, from the 
star to Sgr A*. As a reference, we use IRS 13 E3. Therefore, the ram pressure ratio 
of a given star to IRS 13 E3 is defined as
\begin{equation}
f_{ram} = {\dot M_{wind} \over \dot M_o}{v_{wind} \over v_o}({D_o \over D_*})^2,
\end{equation}
where the quantities $\dot M_o$, $v_o$ and $D_o$ are the wind outflow, wind velocity 
and the distance to Sgr A* of IRS 13 E3. As distance values, $D_*$ and $D_o$ we use 
values given by Rockefeller et al. (2004) obtained assuming that the z coordinate of
 each star is random. Results are given in Table~\ref{tab:1}.

\begin{table}
\centering
\begin{tabular}{c c c}      
\hline 
Source   & $f_{ram1}$ & $f_{ram2}$ \\
\hline 
IRS 16NE &0.11 & 0.027\\
IRS 16NW &0.01 & 0.02\\
IRS 16C  &0.05 & 0.26\\
IRS 16SW &0.22 & 0.62\\
IRS 13E3(AB) &1.0 & 1.0\\
IRS 7W  &0.054 & 0.08  \\
IRS 15SW & 0.011  & 0.02\\
IRS 15NE & 0.018 & 0.02\\
IRS 29N & 0.013 & 0.07\\
IRS 33E &0.1 & 0.02\\
IRS 34W &0.029 & 0.027\\
IRS 16SE &0.014 & 0.03\\
\hline                  
\end{tabular}
\caption{  Results:Values of ram pressure, with the reference source IRS13 E3, for two different values of z coordinate of the source. In the first column the ram pressure was calculated assuming that z coordinate is taken from random distribution, the second column is for z coordinate modeled in Rockefeller et al. (2004).}
\label{tab:1}
\end{table}

\begin{table*}

\centering
\begin{tabular}{c c c c c c c c c c c}      
\hline 
Source   & D  &$V_{wind}$&$V_{star}$& $\lambda_1 [R_{schw}c]$ & ${\cal E}_1/c^2$              &  $\lambda_2 [R_{schw}c]$ & ${\cal E}_2/c^2$ & $\dot{M}_{wind} $& $\dot{M}_{1,accr} $& $\dot{M}_{2,accr}$ \\
  & [arcsec] & [km/s] & [km/s] & &  &  &  & $(10^{-5} M_{\odot} /yr)$& $ (10^{-5} M_{\odot} /yr)$& $ (10^{-5} M_{\odot} /yr)$ \\

\hline  
IRS 16NE & 3.4         & 750      & 560      & 25.0                   & $2.0 \cdot 10^{-6}$ &  16.98                   & $3.99 \cdot 10^{-6}$ & 9.5 & 1.0&0.66\\
IRS 16NW & 8.36        & 750      & 290      & 3.8                    & $1.98 \cdot 10^{-6}$ &  2.57                   & $3.97 \cdot 10^{-6}$ & 5.3 & 0.09 & 0.06\\
IRS 16C  & 4.6         & 650      & 430      & 19.47                    & $1.99 \cdot 10^{-6}$ &  12.14                   & $3.98 \cdot 10^{-6}$&10.5 & 0.94 & 0.59\\
IRS 16SW & 2.8         & 650      & 540      & 54.13                    & $2.04 \cdot 10^{-6}$ &  33.7                   & $4.01 \cdot 10^{-6}$& 15.5 & 3.8 & 2.35\\
IRS 13E3(AB) & 3.8         & 1000     & 207      & 2.16                    & $1.98 \cdot 10^{-6}$ &  1.68                   & $3.97 \cdot 10^{-6}$& 79.1 & 2.6& 1.58\\
IRS 7W   & 8.3         & 1000     & 380      & 1.92                   & $1.98 \cdot 10^{-6}$ &  1.5                   & $3.97 \cdot 10^{-6}$&20.7 & 0.14& 0.1\\
IRS 15SW & 13.4        & 700      & 200      & 1.94                    & $1.98 \cdot 10^{-6}$ &  1.27                   & $ 3.97 \cdot 10^{-6}$& 16.5 & 0.13& 0.08\\
IRS 15NE & 11.5        & 750      & 150      & 1.34                    & $1.98 \cdot 10^{-6}$ &  0.91                   & $3.97 \cdot 10^{-6}$& 18.0 & 0.16 & 0.1\\
IRS 29N & 8.56        & 750      & 220      & 2.71                    & $1.99 \cdot 10^{-6}$ &  1.84                   & $3.97 \cdot 10^{-6}$& 7.3 & 0.12 & 0.08\\
IRS 33E & 3.05        & 750      & 230      & 8.0                    & $1.99 \cdot 10^{-6}$ &  5.43                   & $3.89 \cdot 10^{-6}$& 7.3 & 0.9 & 0.65\\
IRS 34W & 5.8        & 750      & 280      & 5.26                    & $1.99 \cdot 10^{-6}$ &  3.56                   & $3.97 \cdot 10^{-6}$& 7.3 & 0.27 & 0.18\\
IRS 16SE & 8.33        & 750      & 540      & 9.44                    & $1.99 \cdot 10^{-6}$ &  6.4                   & $3.98 \cdot 10^{-6}$&7.3 &0.13 & 0.08\\

\hline                  
\end{tabular}
\caption{Results:$\lambda_1$ and $\dot{M}_{1,accr}$ corresponds to wind temperature
 $T_{wind}=0.5 keV$,  $\lambda_2$ and $\dot{M}_{2,accr}$ to $T_{wind}=1.0 keV$. 
 Mass accretion rate is defined here as $\dot{M}_{accr}=\dot{M}_{wind} \cdot (\frac{R_{BHL}}{D})^2$.}
 Politropic index $\gamma = 1.6$.
\label{tab:2}
\end{table*}

We see that $f_{ram}$ is smaller than 1 for all stars. The highest value, 
obtained for IRS 16 SW is equal to 0.22. For another choice of z coordinates of stars 
(see Rockefeller et al. (2004)), the results are similar. The highest  $f_{ram}$=0.62,
still smaller than 1, is again for IRS 16 SW.

A number of O type stars not listed in Table~\ref{tab:1} also contribute to the
 mass accretion rate, as discussed by Loeb (2004). However, the closest stars have 
too high relative velocity with respect to Sgr A* to allow for mass settlement. For
 example, SO-16 at the closet approach (0.0002 pc) has the velocity of 12 000 km
 s$^{-1}$ and its wind create a narrow stream of gas not directed toward Sgr A*.
 Stars with velocities of 1000 km s$^{-1}$ can lead to an accretion event if they 
are at the distance of 0.015 pc or more. If we consider a star at such a distance
 that its Keplerian velocity is equal to the wind velocity we obtain the maximum 
value of the ram pressure ratio for the star  
\begin{equation}
f_{ram}^{max} = {\dot M_{wind} \over \dot M_o} ({v_{wind} \over v_o} )^3.
\end{equation}
If $\dot M_{wind} \sim 10^{-6} M_{\odot}$yr$^{-1}$ and $v_{wind} \sim 3000 $ km 
s$^{-1}$ occasional episodes of strong accretion from such a star are expected,
in agreement with Loeb (2004) estimates but if $\dot M_{wind} \sim 10^{-6} M_{\odot}$yr$^{-1}$
 and $v_{wind} \sim 3000 $ km s$^{-1}$ then $f_{ram} \sim 0.08$ and 
the wind will be confined by the wind from IRS 13 E3.

In further considerations we assume that the wind of IRS 13 E3 dominates at 
the position of Sgr A*, i.e. it is strong enough to form a bow shock which 
shields Sgr A* from the winds from other stars. In order to check whether this 
assumption  is not in contradiction with the observations we check the 
predicted wind density. The adopted wind outflow rate, wind velocity and the 
distance (see Table~ref{tab:stars}) the wind density at the location of Sgr A* 
(neglecting at this moment the gravitational effect of the black hole) is 108 
cm$^{-3}$.  This value is consistent with the density 130 cm$^{-3}$ 
measured by Baganoff et al. (2003) within the radius 1.4'' of Sgr A*.
 
\subsection{Angular momentum of the flow}

We assume now that IRS 13 E3 is the
dominant source of the matter accreting onto Sgr A$^*$.
We determine the net angular momentum of the flow in the following way:

We consider the case where the 
wind velocity is much higher than the orbital velocity 
of a star. 
Therefore, the material
ejected from a fractional region of the star surface located at $\phi_o$ 
(see Fig.1) 
can reach the gravity center with zero angular momentum (see e.g. Loeb 2004). 
This angle is given by the condition:
\begin{equation}
\sin \phi_o = {v_{star} \over v_{wind}}.
\end{equation}

The wind is mildly supersonic, with the Bondi-Hoyle-Lyttleton
accretion radius, $R_{BHL}$, given by the formula
\begin{equation}
R_{BHL} = {2 G M \over v_{wind}^2 + v_s^2},
\end{equation}
where $v_{wind}$ is the flow velocity and $v_s$ is the sound speed 
within the wind.

The spherically-symmetric wind blowing at $\phi \ne \phi_o$ or out of the orbital 
plane will posses some
amount of positive or negative specific angular momentum
(angular momentum density), $l$. In the second order
approximation 
\begin{equation}
l \approx -v_{wind} D \delta \phi \cos(\phi_o) [1 - 1/2 \tan \phi_o \delta \phi],
\label{eq:l_loc}
\end{equation}  
where $\delta \phi = \phi - \phi_o$, $\phi$ is the azimuthal angle of the element at the
star surface and $D$ is the distance between the star and the
Galactic center. 

A cylindrical fraction of this flow, with $\Delta \phi = R_{BHL}/D$, and $\Delta \theta
\sim R_{BHL}/D$ will be intercepted by the central black hole, 
where the angle $\theta$ determines the deviation from the orbital plane. 
Integrating the Equation~\ref{eq:l_loc} with respect to 
$\delta \phi $ and $\delta \theta$ in
the limits specified by $\Delta \phi$ and $\Delta \theta$,
 we obtain the net angular
momentum of the flow as:
\begin{equation}
l_{eff} = {2 \over 3 \pi} {1 \over \sqrt{1 - (v_{star}/v_{wind})^2}} v_{star}D 
\left({R_{BHL} \over D}\right)^2,
\label{eq:leffzero}
\end{equation}
The above relation is 
valid if the star velocity is significantly smaller than the wind velocity. Large value of
the angular momentum density of the donor star, $v_{star}D$, is decreased by small 
quadratic term in the $R_{BHL}/D$ ratio.

We further express the angular momentum density in dimensionless units 
$2 GM/c$, or equivalently
\begin{equation}
\lambda = {l_{eff} \over R_{Schw} c},
\end{equation}
where $R_{Schw} = 2GM/c^2$.

We calculate the values of $\lambda$ for each of the stars in 
Table~\ref{tab:1}. We consider two values of the temperature representative
for the possible wind temperature: 0.5 keV and 1.0 keV (eg. Muno et al.2004, Raassen et al. 2003). 
 We calculate angular momentum $\lambda$ for these two values of wind temperature , $\lambda_1$ 
(also Bernoulii constant ${\cal E}_1$) corresponds 
to 0.5 keV , $\lambda_2$ (and ${\cal E}_2$) to 1.0 keV.

In the case of the temperature of 0.5 keV the average value of 
$\lambda$ calculated for all stars is 12.2, and if we neglect the dominant star 
IRS 13 E3 this mean value is equal to 13.3. It is interesting to compare 
this value with the mean angular momentum obtained in hydrodynamical 
simulations in the innermost part of the flow. In dimensionless units, these 
values are $\sim 25$ in Cuadra et al.(2005b), $\sim 40-60$ in Coker $\&$ Melia (1998),
 and 3-20 in Coker $\&$ Melia (1997),
 not far from our results. However, as we showed above (see Sect.~\ref{sect:winds}) actually the
star IRS 13 E3 dominates the flow and the effective angular momentum is 
ten times lower. It shows that using equally efficient stellar winds in the 
simulations may lead to unrealistic description of the flow.

\subsection{Flow geometry within $R_{BHL}$ and the Bernoulli constant}

The supersonic flow develops a shock at a distance to the black hole 
comparable to $R_{BHL}$, as shown in a number of papers addressing the issue
of the supersonic motion of the accreting objects (e.g. Salpeter 1964, Bisnovatyi-Kogan 1979). 
The topology of solutions was systematically discussed by number authors (Hunt 1971, 1979, 
Livio et al. 1979, Okuda 1983, Livio et al. 1986, Shima et al.1986,
 Matsuda 1992, Ishii et al. 1993 Ruffert 1994,1995,1996)
Three dimensional calculations recently were made by Pogorelov et al. (2000), and
Foglizzo et al. (2005).

Since black hole has no rigid surface the shock does not initially form in 
front of it but only sideways, as sketched in Fig.~\ref{fig:draw1}. Shocked
material returns toward the black hole and the dominant inflow actually comes 
from this back-side flow. Since the infalling material has certain small 
angular momentum and the density slowly decreasing perpendicularly to the flow
in the same direction as the donor star velocity the flow patters is not 
symmetric with respect to the main axis. The stagnation point is shifted, 
and the material with the lowest and the highest angular momentum mix. 

We can determine the Bernoulli constant of this material,  $\cal E$, either
above or below the shock. Its
value above the shock can be estimated as 
\begin{equation}
{\cal E} = {1 \over 2}v_{wind}^2 + {v_{s,wind}^2 \over \gamma_i - 1}-
 {GM \over (R_{BHL}-R_{Schw})} + {1 \over 2} {\lambda^2 \over R_{BHL}^2 }
\label{eq:Bern1} 
\end{equation}
assuming that the velocity of the wind close to the shock is equal to initial wind velocity, 
and that the shock occurs at distance comparable to $R_{BHL}$. In Eq.~\ref{eq:Bern1} the second term
dominates, so the Bernoulli constant is mainly determined by assumed wind temperature $T_{wind}$ and politropic 
index $\gamma$.

Values of $\cal E$, in $c^2$ units, are given in Table~\ref{tab:1}, for two values 
of the wind temperature.  Because for a supersonic wind the Bernoulli constant
 $\cal E$ is $\sim v_{s,wind}^2/v_{wind}^2 
\times \frac{3-\gamma}{2(\gamma-1)}$ all values of $\cal E$ in the Table~\ref{tab:2} are almost the same for 
corresponding temperature of the wind.

 For IRS 13 E3 they are of order of  $(1 - 4) \times 
10^{-6}$.
In calculations we adopt $\lambda$=2.16 and $\cal E$ for $T=0.5$ keV 
before the shock as appropriate for consideration of the 
accretion close to Sgr A*.
Because we assume adiabatic shocks, 
the Bernoulli constant doesn't change after passing the shock. 
Keeping the Bernoulli constant we obtain 
temperature of the gas after a shock to be $T= 1.4 keV$ (for parameters adopted for Sgr A*).
This value seems to be in a good agreement with X-ray observations of Sgr A* (Baganoff et al. 2003).

In order to simplify the issue we further assume that the accretion proceeds
roughly in a spherically symmetric way, with the effective angular momentum 
given by Eq.~\ref{eq:leffzero} and the Bernoulli constant determined from
Eq.~\ref{eq:Bern1}. This simplification allows to use the semi-analytical 
solution for the flow pattern outlined in the next section.

\section{Low angular momentum flow close to Sgr~A*}

Therefore, we further study the low angular momentum 
(strongly sub-Keplerian) flow 
onto Sgr A$^*$. Apart from Cuadra et al. (2005a,b), only a few 
papers were devoted to this option so far. Melia
(1994) considered a model of a free fall with non-negligible angular
momentum assuming no contribution to emission from within the 
circularization radius. Proga \& Begelman (2003a,b) performed
2-D hydrodynamical and magnetohydrodynamical simulations of the
low angular momentum flow for a specific class of outer boundary 
conditions (angular momentum decreasing with the distance from the 
equatorial plane).
 
Here we concentrate on semi-analytical solutions with standing shocks
in the pseudo-Newtonian potential.

\subsection{General character of the low angular momentum flow}

The initially subsonic flow onto a black hole must exhibit
transonic behavior in order to satisfy the inner boundary 
conditions imposed by the
event horizon, i.e. it becomes supersonic again.

Low angular momentum flow may actually posses more than one sonic, 
as first shown
by Abramowicz \& \. Zurek (1981). Typically the external
sonic point, $r_{out}$, lies far from the black hole 
(close to a Bondi radius for a corresponding spherical accretion).
The internal sonic point, $r_{in}$, and the middle 
sonic point $r_{mid}$ exist within and outside the marginally 
stable orbit, respectively, for general relativistic (Das 2004,
Barai, Das \& Wiita 2004) as well as for post-Newtonian 
(D02, DPM) model of accretion flow.
The location of the sonic points can be calculated as a function
of the specific flow energy ${\cal E}$ (the Bernoulli's constant),
angular momentum $\lambda$ and inflow polytropic index $\gamma_i$
(see, e.g. \S 3 of D02 for details of such calculations).

If $\lambda$ is close to zero, a shock does not form, and accretion 
remains supersonic down to the event horizon after it crosses $r_{out}$.
For slightly larger $\lambda$ the centrifugal barrier 
becomes strong enough, inflowing matter starts pilling up 
close to the black hole due to the resistance offered by 
the barrier, and the depleted matter may break the incoming 
flow behind it and consequently a shock forms. Such shocks 
may become steady and standing so that they can be studied
within the framework of stationary flow (D02, DPM and references
therein). 

Following D02, we consider here a stationary non 
self-gravitating, non-magnetized, inviscid accretion of polytropic fluid. 
We assume that the flow proceeds through a standing shock, so the 
discontinuity in the radial velocity allows to match the supersonic flow 
below $r_{out}$ with the transonic flow through $r_{in}$. The exact 
location of the shock as a function of parameters
$\left[{\cal E},\lambda,\gamma_i\right]$ is obtained by solving the 
generalized Rankine-Hugoniot equations. We assume that the shock is 
non-radiating and infinitesimally thin.

However, for the parameters chosen as the best values for the accretion flow onto
Sgr A* (${\cal E}=2 \times 10^{-6} - 4 \times 10^{-6}$ , $\lambda = 1.68 - 2.16$ and
 $\gamma_i = 1.6$, see results for IRS13E3 in Table~1)
we do not obtain a stationary solution describing the accretion all
the way down to the black hole. 

In order to understand what is happening we analyze the model properties
in the broader parameter space around the adopted values.

\subsection{Dependence of the flow topology on model parameters}

\begin{figure}  
\epsfxsize = 85 mm  
\epsfbox{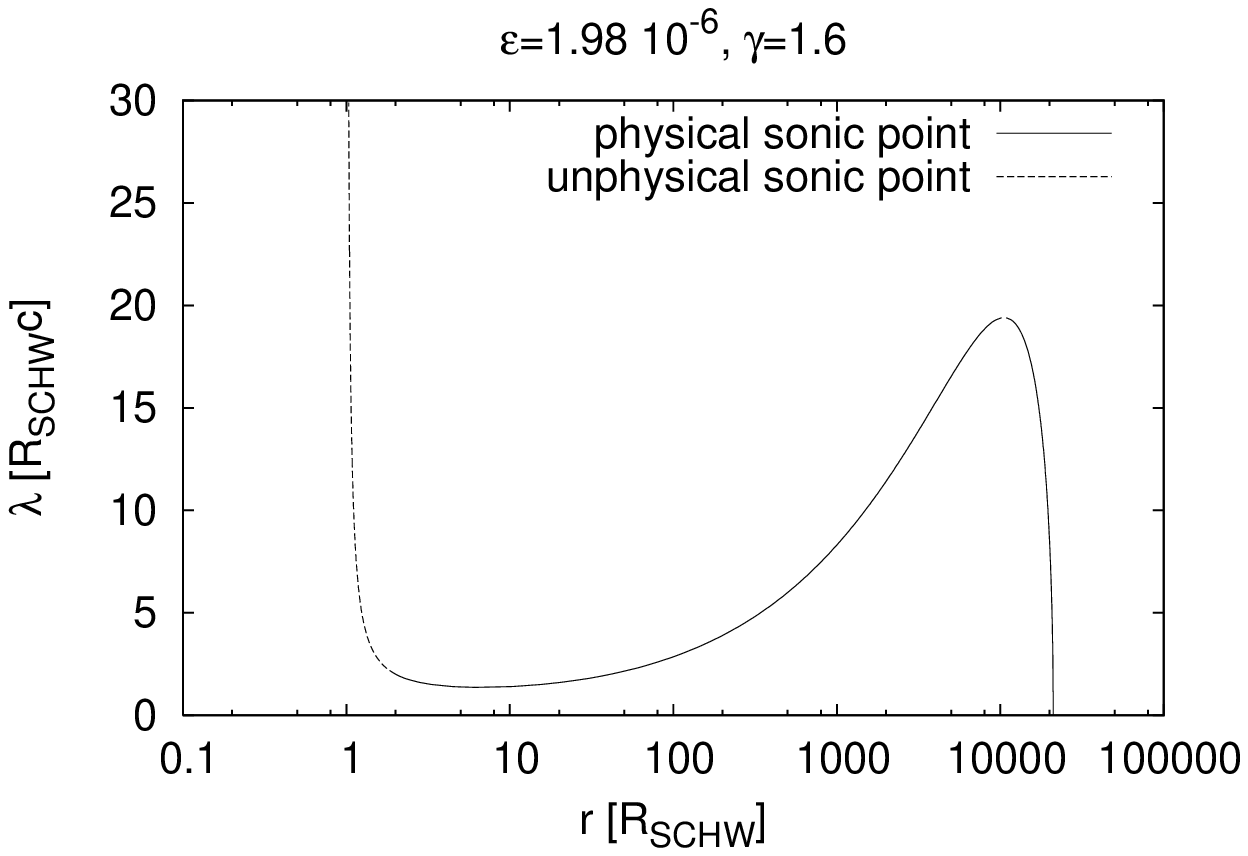}
\epsfxsize = 85 mm  
\epsfbox{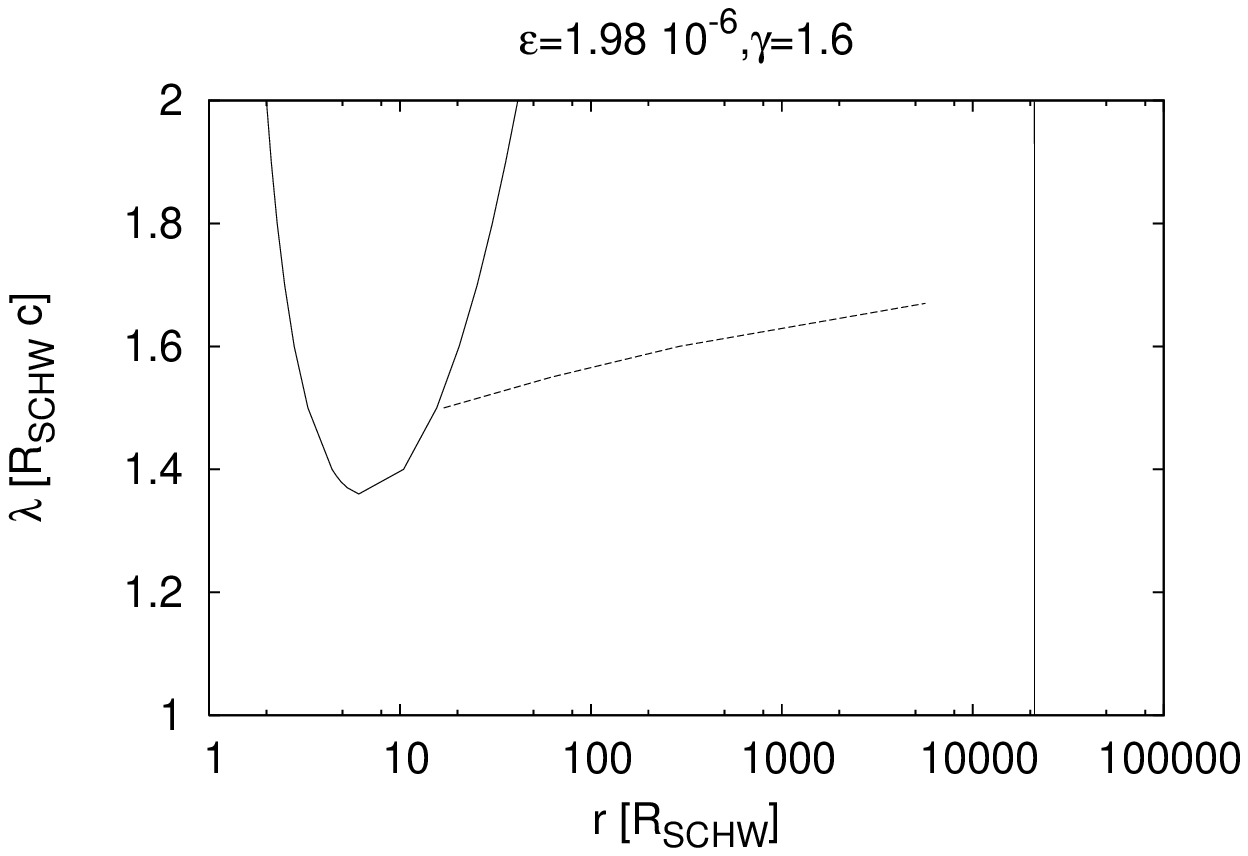}

\caption{The position of the sonic points as functions of the angular 
momentum for the Bernoulli constant appropriate for Sgr A* accretion flow. Lower panel
shows the expanded version of the upper graph, with the position of the shock
marked with the dashed line. For $\lambda = 2.16$ no stationary shock solution exists.
\label{fig:sonic1}}  
\end{figure}

For a fixed $\gamma_i$ and ${\cal E}$ the position of the sonic point and the
position of the shock are strong functions of $\lambda$. 
The equation of motion with small angular momentum allow us to determine the position of the sonic points: 
\begin{equation}
{\rm{v^2} \over 2}+{\rm{v_s^2} \over \gamma-1 } - {GM \over r - R_g } + 
{\lambda^2 \over 2 r^2}= \cal E
\label{eq:motion}
\end{equation}
 Derivative of Eq.\ref{eq:motion} can be expressed as:
\begin{equation}
\rm{v {dv \over dr}} =\frac{{2\over r} \rm{v^2} - {GM \over (r-R_g)^2} + 
{\lambda^2\over r^3}}{1-{v^2\over v_s^2}},
\label{eq:motion_integral}
\end{equation}
 with the help of the continuity equation.
The sonic point condition is that nominator and denominator of Eq.\ref{eq:motion_integral}
 equals zero. In the upper panel of Fig.~\ref{fig:sonic1} we show the position of sonic points, as  
functions of $\lambda$ parameter for constant Bernoulli constant 
${\cal E} = 1.98 \times 10^{-6}$ in $c^2$ units.

For $\lambda$ lower than about 1.4 (in general for any Bernoulli constant), one 
solution of equations Eq.\ref{eq:motion} and Eq.\ref{eq:motion_integral} exist above
the black hole horizon for introduced
gravitational potential. 
This point is located at a large distance from
the black hole and the flow proceeds 
as in the case of a spherical Bondi flow. 
For $\lambda$ between $\sim 1.4$ and $\sim 20$ (in general upper value of 
$\lambda$ 
is a function of specific flow energy $\cal E$) three formal solutions exist.
However, there is a significant difference between the solutions for  $\lambda$ 
smaller than 2 and larger than 2. For $\lambda$ smaller than $\sim 2.0$ (exact limit
slightly depends on $\cal E$) the inner sonic point describes the physically
acceptable solution for the inflow.  
However, for $\lambda$ larger than $\sim 2.0$ the internal sonic 
point is unphysical, in this point
$v_s^2$ is less than zero (see Abramowicz \& \. Zurek 1981 for the discussion).
The transition from physical to unphysical inner sonic points takes place for $\lambda$
very close to 2 (the dependence on $\cal E$ is of order of $\cal E$ in 
our dimensionless units.
In Fig.~\ref{fig:sonic1} (upper panel) we mark the unphysical points with dashed line,
 and the physical points with solid line.

For $\lambda$ still larger than $\sim 20$ again one sonic point exists (unphysical ones), 
it is located under the last stable orbit (sonic point position approach to 1 $R_g$)
and the flow proceeds in a subsonic way.
This type of solution finally becomes unphysical since in the case of higher angular
momentum subsonic flow it is unlikely that the angular momentum transfer is negligible
and the appropriate solution of the physical flow is provided by ADAF-type models.

In the lower panel of Fig.~\ref{fig:sonic1} we expand the figure around most
plausible values of $\lambda$ for Sgr A*, and we mark the position of the stationary 
shock with the dashed line. We see that this type of solutions exist in a very
narrow range of the physical parameters, as noted before in D02. 

\begin{figure}  
\epsfxsize = 85 mm  
\epsfbox{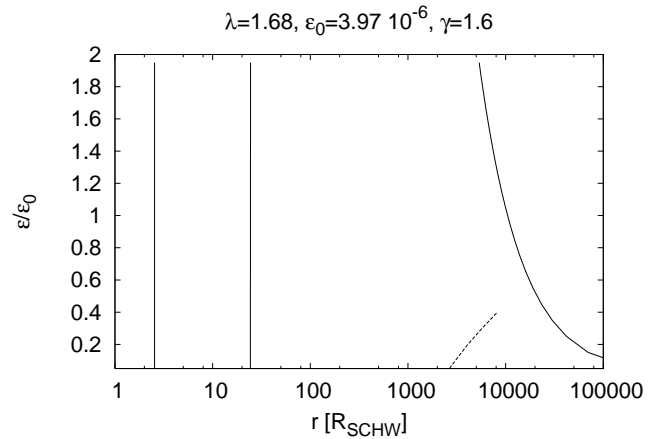} 
\caption{The position of the sonic points as functions of the Bernoulli 
constant for angular momentum  $\lambda=1.68$ appropriate for Sgr A* accretion flow. Position of the shock
is marked with the dashed line.
\label{fig:sonic3}}  
\end{figure}

Fig.~\ref{fig:sonic1} was obtained for a specific value of the Bernoulli constant
but the dependence on it is not very strong. In Fig.~\ref{fig:sonic3} we
show the positions of the sonic points as functions of $\cal E$ for fixed $\lambda$,
for higher initial wind temperature 1.0 keV. 
Stationary shock develops only for the Bernoulli constant a few times lower that
the value estimated for Sgr A*. The same picture is not possible for $\lambda=2.16$
because in the case of $\lambda > 2$ the inner sonic points are unphysical, so the
transonic accretion for lower specific energy of the flow 
is still not possible. Changing of $\cal E$ is changing only the position 
of the outer sonic point (as shown in Fig.~\ref{fig:sonic3}), so in a case of $\lambda > 2$ 
with lower $\cal E$ the topology will not change.

\begin{figure}  
\epsfxsize = 85 mm  
\epsfbox{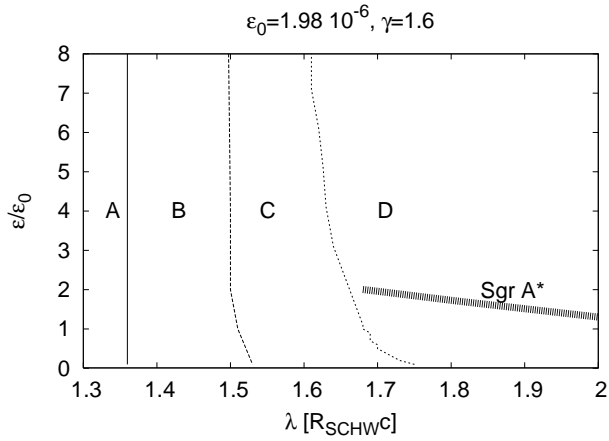} 
\caption{  Region A corresponds to semi-spherical flow without a shock
and with a distant sonic point and inner under horizon,
in region B four sonic points exist, the condition for a stationary shock is
never satisfied but the flow proceeds transonically without a shock. In region C
a stationary shock solution exists, D region corresponds to the parameters for
 which torus form. Plausible parameters of the inflow in Sgr A* locate the
solutions in region D (thick continuous line). 
\label{fig:sonic4}}  
\end{figure}

We can mark conveniently the types of solutions using the $\cal E$ - $\lambda$
diagram (see Fig.~\ref{fig:sonic4}), as introduced in D02. We have four regions in 
the interesting range
of parameters for Sgr A*. Region A corresponds to semi-spherical flow without a shock
and when we have only one distant sonic point.
In region B three sonic points exist, the condition for a stationary shock is
never satisfied but the flow proceeds supersonically without a shock. 

\begin{figure}  
\epsfxsize = 85 mm  
\epsfbox{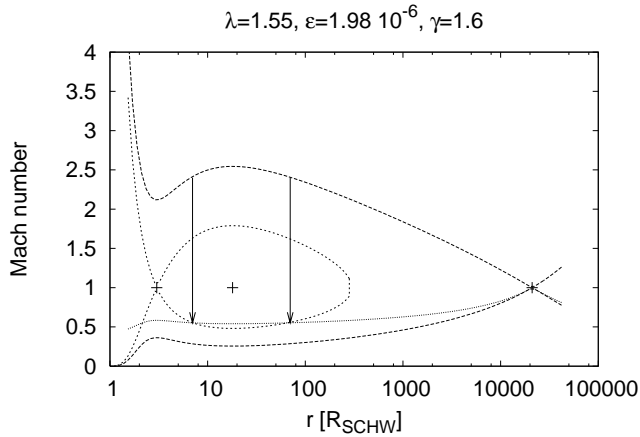} 
\caption{The topology of the flow obtained wit a standard stationary
shock solution for angular momentum lower than that for 
Sgr A*. Dotted line marks the position of the line
where Rankine-Hugoniot conditions are satisfied, so the intersection of this line 
with the transonic flow lines marks the position of the shock (arrows). Inner 
shock is unstable, outer shock is stable. Sonic points are marked with pluses '+'.  
\label{fig:topo2}}  
\end{figure}

Fig.~\ref{fig:topo2} shows the topology of the shocked accretion flow for
the values of the Bernoulli constant $\cal E$ and $\gamma_i$ appropriate 
for Sgr A*, but with angular momentum $\lambda=1.55$, lower than the best estimate
for Sgr A*. Such a topology is representative for region C.
For such $\lambda$ the stationary 
shock is possible. Matter first passes through 
$r_{out}$ ( $\sim 10^4$ $r_g$) and
encounters a shock at $r_{sh}$ ($\sim$ 70 $r_g$) close to the black hole.
The solid vertical line marked with a down-ward arrow represents the 
shock transition.
$M_-$ and $M_+$ are the pre/post shock Mach numbers and the shock strength 
${\cal S}$ is defined as ${\cal S} = \frac {M_-}{M_+}$, which comes out to be
4.0 for this case. Post-shock subsonic
inflow becomes supersonic again after crossing $r_{in}$ ( 3.1 $r_g$) and
finally dives through the event horizon.

\begin{figure}  
\epsfxsize = 85 mm  
\epsfbox{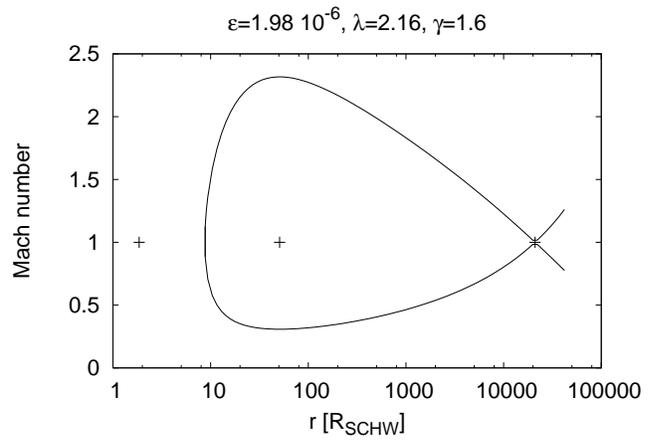} 
\epsfxsize = 85 mm  
\epsfbox{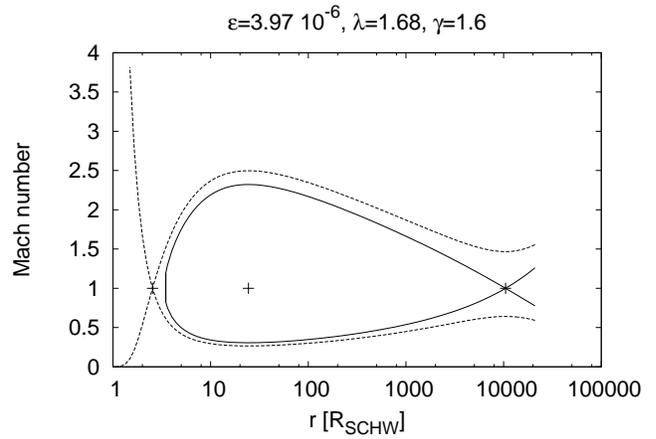} 
\caption{The topology of the flow obtained for the flow 
parameters corresponding to Sgr A*. Close loop touches the radius
 of 4 $R_{Schw}$ in a case of $\lambda=1.68$
or 10  $R_{Schw}$ in a case of $\lambda=2.16$
and the constant angular momentum torus forms there.Sonic points marked with pluses '+'.
\label{fig:topo1}}  
\end{figure}

Sgr A*, however, lies in the region D, according to our estimates.
In the region D, three sonic points still exist but the
topology is different, the initially supersonic flow proceeds along the
flow line which end up at a distance of few Schwarzschild radii,
since the flow line forms there a closed loop (see Fig.~\ref{fig:topo1}). 
The angular momentum barrier is just strong enough to prevent the flow 

This dependence of the topology on the flow parameters was
discussed before (see D02, Das 2004, Chakrabarti \& Das 2004). 
However, most of the attention in those papers
was payed to models with stationary shocks, located in region C. 
What is new and important is that
our best estimates of the flow in Sgr A* locate the flow in the D region, where
no stationary shock develops.

This means that the flow in Sgr A* is non-stationary in a natural way, and it
consists of two parts:
\begin{itemize}
\item  outer, semi-stationary flow  
\item inner, non-stationary torus
\end{itemize}

\subsection{Constraints on mass accretion rate from Faraday rotation in an outer flow}
Our model gives self-consistent constraints on the mass accretion rate, which is
 estimated from the conditions at the position of the first outer sonic point.
The values of mass accretion rate estimated by this condition are given in Table~\ref{tab:3}.
In Table~\ref{tab:2} we showed the accretion rate calculated using the formula:
$\dot{M}_{accr}=\dot{M}_{wind} \cdot (\frac{R_{BHL}}{D})^2$, which is about an order of magnitude
larger than the one calculated at 
the position of the sonic point. Thus we assume here that there must be an outflow at some radius,
 but we do not specify its form or geometry.
\begin{table}
\centering
\begin{tabular}{c c c}      
\hline 
$\lambda$ & $\dot{M}_{sonic point} [M_{\odot} /yr]$ & $RM [rad/m^2]$\\
\hline 
0.82 & $3.93 \cdot 10^{-6}$ & $5.7 \cdot 10^5$\\
1.68 &$ 2.45 \cdot 10^{-6}$ & $7 \cdot 10^5$\\
2.16 &$ 2.18 \cdot 10^{-6}$ & $2 \cdot 10^6$ \\
12.2 &$ 2.15 \cdot 10^{-6}$ & $2.8 \cdot 10^6$ \\
\hline                   
\end{tabular}
\caption{ Mass accretion rate for four values of angular momentum $\lambda$ parameter.}
\label{tab:3}
\end{table}
 
We compare our results with the limitations estimated from Faraday rotation measurements, which give us 
a limit for the mass accretion rate very close to the black hole (up to 100 $R_{SCHW}$).
In general such high accretion rates as shown in Table~\ref{tab:3} 
give RM coefficient to be very high in comparison to the data.
For non relativistic plasma the rotation measured is given by:
\begin{equation}
RM=8.1 \cdot 10^5 \int n_e \vec{B} d \vec{l}
\end{equation}
For ultra-relativistic thermal plasma ($Te > 6.0 \cdot 10^9$) RM have to be additionally 
multiplied by a factor of $\frac{log \gamma}{ 2 \gamma^2}$ (Quataert $\&$ Gruzinov 2000).
Taking into account our $n_e(r)$ profile and the absolute value of the 
magnetic filed (which is in equipartition with the gas) we obtain
RM to be overestimated.
However, the magnetic field of the inflowing material is unlikely to have large scale structure. 
Indeed, from magnetohydrodynamical simulations of 
accretion of plasma with low angular momentum we know 
that if we include changing of the direction of the magnetic field along any specific line of sight the
RM factor is reduced even a few hundred times (Moscibrodzka et al. 2006, in preparation). Assuming 
the reduction factor of 200 the results, given in Table~\ref{tab:3},
are consistent with the recent measurements of RM for Galactic center where
RM is from few times $10^5 rad/m^2$ to few times 
$10^6 rad/m^2.$ (Bower et al. 2005, Quataert $\&$ Gruzinov 2000).

\subsection{Stationary emission from the outer part of inner flow}

The outer part of the inner flow, from the Bondi radius down to the 
inner radius of the closed loop in
Fig.~\ref{fig:topo1} can be considered approximately as a stationary. Knowing
the flow parameters, we can calculate the radiation spectrum of such a flow,
and we expect it to represent the stationary, non-variable emission
of Sgr A*.

We use the model outlined in Mo\' scibrodzka et al. (2005). This model is
based on the assumption that the flow is two-temperature, with the flow 
dynamics mostly determined by ions. Electrons are heated partially due to
the Coulomb coupling with ions, but also $\delta$ fraction of energy goes
directly to electrons through the Ohmic heating.
The presence of the magnetic field is assumed, with the field
strength measured with respect to the equipartition value through the $\beta$
parameter. The emission is calculated with the use of the Monte Carlo code,
taking into account bremsstrahlung, synchrotron emission and Comptonization.
 
\begin{figure*}  
\epsfxsize = 170 mm  
\epsfbox{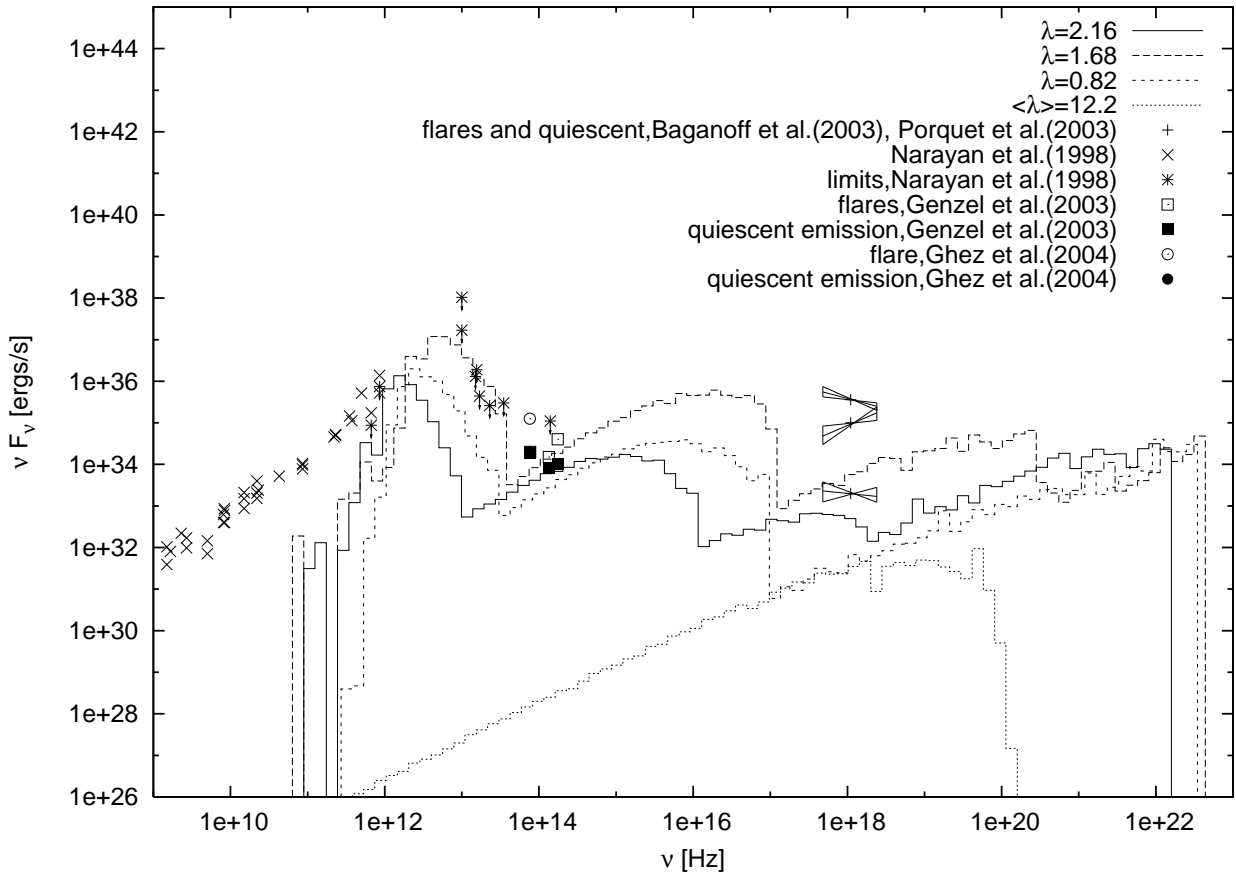} 
\caption{The stationary emission of the Sgr A* flow generated between the outer
radius, which is about 2 Bondi radii and the inner radius determined by the 
distance of the closed
loop flow from the black hole. For $\lambda=1.68$ the inner loop close at
 distance 4.0 $R_{Schw}$ , for $\lambda=2.16$ the loop close at about 10 $R_{Schw}$ 
and spectrum is marked as long dashed and solid line respectively.
The spectrum marked as dotted line is calculated for the mean value 
of specific angular momentum $<\lambda>=12.2$ of all available sources of matter,
in this case the torus forms at about 1000 $R_g$. For $\lambda=0.82$ (the case of 
IRS 13 but with the assumption
that the star cluster is located 10 arcsec from the center), the flow is continuous 
down to the event horizon 
without formation of a shock and the spectrum is marked as short dashed line.
 \label{fig:stationary_spectrum}}  
\end{figure*}
Models in Mo\' scibrodzka et al. (2005) were based on spherical accretion 
dynamics but now we generalized to the case of the flow with low angular momentum,
described
in Section 3.2. The adopted parameters for modeling emissivity are $\beta =
0.5$ and $\delta = 0.5$.

The spectra calculated for four representative values of the angular
momentum
are shown in Fig.~\ref{fig:stationary_spectrum}.

The most satisfactory representation of the broad band stationary emission
is
provided by the model with $\lambda = 2.16$. It fits the quiescence NIR
spectrum quite well and does not overpredict the stationary X-ray emission
measured by Chandra. The spectral slope of the modeled X-ray emission is
within the
error bars of the data. The radio tail is not reconstructed. To fit radio
part
of the spectrum, we need a non-thermal population of electrons
(Yuan et al. 2003), currently not included in the model.
 The model with $\lambda = 1.68$ is slightly too bright
in the X-rays and in the IR. This is due to the fact that the inner boundary
of a stationary flow is further out for $\lambda = 2.16$ and closer in for
$\lambda = 1.68$.

Since our determination of the most plausible range of angular momentum is
the
accreting flow is based on a number of assumptions, we also show the
resulting
spectrum for a lower, and a higher value of $\lambda$.

The solution $\lambda = 0.82 $ corresponds to a distance of IRS13 of 10
arcsec.
Such a solution has different topology, as seen from Fig.~4. It belongs to
the
region A, which means that a single (outer) sonic point exists, and the flow
is
practically as in the case of a flow without angular momentum. Flow velocity
is
high and the density is relatively low although the accretion rate required
by
the condition of transonic solution is actually slightly higher than in the
two cases considered before. Therefore, the radiative efficiency of the flow
becomes lower although the stationary flow proceeds down to the black hole
horizon. The resulting spectrum (see Fig.~7) underrepresent the X-ray
persistent emission by more than an order of magnitude.

The solution  $\lambda = 12.2 $ corresponds to the assumption that the ram
pressure cannot prevent accretion streams from various stars to approach Sgr
A*
(i.e. the value of the angular momentum is given as a mean value for all
stars
from Table~2). This solution belongs to region D in Fig.~4. The high value
of
the angular momentum prevents the stationary flow from approaching a black
hole
 - the accumulation radius is $\sim 1000 R_{Schw}$. The level of emission is
very low and the spectrum and none of the observational points are
reproduced.
It may not conclusively rule out all higher angular momentum models since
the
model of constant angular momentum flow is not likely to apply to the flow
with
high initial angular momentum. Effects of the angular momentum transfer may
in such case take place, as suggested by the numerical simulations (e.g.
Cuadra et al.2005). The issue can be set by future
MHD simulations with appropriate grid coverage, realistic boundary
conditions
and computations of the resulting spectra.

\subsection{Inner torus}

At the inner radius of the semi-stationary flow, 
the inflowing matter accumulates and forms a ring
of material with constant angular momentum. The exact location of the 
torus/ring depends on the angular momentum. For $\lambda = 1.68$ it forms
very close to the marginally stable orbit. For $\lambda = 2.16$ (determination
for IRS 13 E3 assuming lower wind temperature; see Table~\ref{tab:1}) the
topology of the solution is again as in Fig.~\ref{fig:topo1} but the 
ring forms at 10 $R_{Schw}$.

The equilibrium of the material with fixed angular momentum close to a 
black hole
has been studied long time ago (Abramowicz, Jaroszy\' nski \& Sikora
1978, Jaroszy\' nski, Abramowicz \& Paczy\' nski 1980). The key parameter 
here is the exact
value of $\lambda$. If $\lambda < 2$ then the equipotential surface exists
with an inner cusp. As soon as the material fills this equipotential surface 
the accretion proceeds. This means that the inflow is completely rebuilt and
the initial mostly supersonic inflow is replaced with the solution which is 
subsonic above the marginally stable orbit. It is represented by the lower line
in Fig.~\ref{fig:topo1}. Such a solution, if reached, would
remain stable. However, the timescale to built this flow pattern is long. For
our $\lambda = 1.68$ solution the mass in the supersonic branch (upper transonic
solution in Fig.~\ref{fig:topo1}) is  $ M_1 = 1.28 \times 10^{-5} M_{\odot}$, 
the mass in the subsonic branch (the lowest branch in Fig.~\ref{fig:topo1} in lower panel) is 
$M_2 = 1.22 \times 10^{-5} M_{\odot}$,
and the accretion rate on the transonic branch is 
$\dot M = 2.45 \times 10^{-6} M_{\odot} yr^{-1}$, so the reconstruction 
timescale is roughly $(M_2 - M_1)/\dot M = 0.26$ yr. 

On the other hand, the newly formed ring is known to be strongly unstable, 
roughly in the local
dynamical timescale (Papaloizou \& Pringle 1984). Actually, a ring model as an
explanation of the Sgr A* variability (e.g. Liu \& Melia 2002, Rockefeller et al 2005, 
Prescher 2005, Tagger \& Melia 2006).
Therefore, we find it extremely interesting that the best estimates of the
accretion flow parameters leads in a natural way to formation of such a ring.

Calculation of the dynamics and the spectra of this ring is beyond the scope
of the present paper. We can only speculate, that the matter in a ring 
accumulates as long as the density contrast between the ring and the infalling
material is large enough. In that case the (non-stationary) shock forming 
between the ring and the inflowing material can effectively reflect the 
instability waves and the torus is violently disrupted. Most of the material
plunges into black hole or is ejected in a form of a jet. However, small 
fraction of the material may remain in a ring-like configuration, with 
increased angular momentum. Hydrodynamical simulations (\. Zurek \& Benz 1986, 
Hawley 1987; see also Proga \& Begelman 2003ab and the references therein) and analytic
studies in the non-linear regime (Goodman, Narayan \& Goldreich 1987) support
the instability scenario but the long time-scale behavior of the flow under
such conditions is not determined yet. The magnetic field is also likely to play
a role in redistribution of the angular momentum (e.g. De Villiers \& Hawley 2003,
Tagger \& Melia 2005) or producing an outflow.

\section{Discussion}

The accretion flow into Sgr A* black hole is widely believed to be 
radiatively inefficient but the
exact character of the inflow is under discussion. 
In the present paper we estimated analytically
the effective angular momentum of the accretion flow by analyzing the relative strength
of the winds of the nearby stars.

 We calculated ram pressure from each of the sources listed in Table 2.
 Assuming the stellar wind properties, positions and velocities
given in Rockefeller et al. (2004) we argue
 that IRS 13E3 dominates. This star complex would stop dominate if it was
more than 5 arcsec far away from the center. 
If IRS 13 is further away IRS16 SW would be a dominating source instead,
because it has second high mass accretion rate.
In such case the angular momentum supplied by this star complex would be too high to 
accrete directly
closer than $10^6 R_{SCHW}$. An angular momentum transfer would be requested
to power Sgr A* and the dynamics would be like in ADAF flow.
 On the other hand if we slow down the wind of IRS 13 has a velocity of 650 km/s (the same value
 as of IRS 16SW), with the distance unchanged, then IRS13 still dominates. 

Therefore, we favor the scenario in which the wind from the IRS 13 E3 complex 
dominates, and the angular momentum, $\lambda$, is of order of 1.68 - 2.16 in dimensionless
units ($R_{Schw}c$). (Angular momentum provided by IRS 13E3 can change due to
 ellipticity of the orbit ,and the picture of accretion 
flow can change. The issue concerning ellipticity of orbits was discussed in context
of very near stars with short periods (Loeb 2004), but it is not discussed in this paper.)
. This value is much lower than usually adopted in the outer part of
the ADAF-type flow. Thus we analyzed the flow in the pseudo-Newtonian potential 
following the original ideas of Abramowicz \& \. Zurek 
(1981) (see Das 2004 for recent developments) allowing for a multi-transonic solution 
for constant angular momentum. We found that the angular
momentum estimated for Sgr A* is just large enough to provide the centrifugal barrier
close to the black hole. Therefore, the inflow is mostly supersonic but close to the black
hole the stationary solution does not exist and the material piles up. 
The radius, where an inner ring/torus forms depends significantly
on the exact value of the angular momentum, and for $\lambda$ between 1.68 - 2.16 it falls
into 4 - 10 $R_{Schw}$ range. We assumed that the inflow region is the source
of the persistent emission and we calculated the radiation spectrum for the two extreme
cases of $\lambda$ equal to 1.68 and 2.16, taking into account bremsstrahlung, synchrotron 
emission and Comptonization. The accretion rate ($\dot M = 4.29 \times 10^{-6} 
M_{\odot} yr^{-1}$) was determined self-consistently from
the model, from the condition of the transonic inflow in the outer region. 
Taken at the face value, this accretion rate overpredicts the observed
Faraday Rotation measure by two orders of magnitude (Bower 2003,2005). However, we
find that if magnetic field is very tangled, then this accretion rate 
does not violate this limits.

The case of larger $\lambda$ gave the spectrum roughly 
consistent with the persistent emission, the case of smaller $\lambda$ overpredicted the
persistent emission level. 

The inner ring/torus was not modeled in our paper, but is is well known that such a
constant angular momentum ring is a subject to violent instabilities (Papaloizou \& Pringle
1984)
which opens an attractive scenario for modeling SGR A* flares. Therefore, our estimates of
the inflowing material parameters support the models based on the presence of an inner torus
(e.g. Liu \& Melia 2002, Rockefeller et al. 2005, 
Prescher 2005, Tagger \& Melia 2006).

Our analytical estimates are necessarily oversimplified, and it would be interesting to see
whether 3-D computations, like those of Cuadra et al. (2005a,b) give the same result if the
dominant role of IRS 13E3 is taken into account. However, this is not simple since it 
requires modeling also in the supersonic region. 

If our estimates of the angular momentum and the accretion rate are correct then we deal with
two issues important for the long timescale studies. 

First issue concerns the accretion rate secular changes. In our model the wind accretion
rate given in Table~1 is by a factor of 3 - 5 higher than the accretion rate through the
outer sonic point. This means that some material is temporarily stacked there, at distance
of a few thousands $R_{Schw}$. We estimated the timescale for the radiative cooling in
this region from our model and this timescale is long ($2.8 \times 10^6$ yr) 
which means that the material 
remains hot. Second issue is the cumulating angular momentum of the material in the 
innermost part of the flow. The inner ring/torus instabilities cause events of accretion but
a fraction of the material has to remain to carry the excess angular momentum and a kind of
equatorial outflow/drift of high angular momentum material is likely to develop. Therefore,
we can expect some kind of instability in a longer timescale, when this accumulating 
material finally forms a type of an ADAF inflow with much higher accretion rate. Such
periods of enhanced accretion probably occasionally happen in Sgr A* (for evidences for an
enhanced luminosity about 300 yr ago, see Revnivtsev et al. 2004). 
Unfortunately, performed numerical simulations do not cover yet timescales and radial ranges 
wide enough to see such phenomenon.   

\section{Conclusions}

Our analysis of the flow pattern in Sgr A* indicates that

\begin{itemize}
\item If IRS13 distance to Sgr A* in 3D
is close to its projected distance of about 3.5,
IRS 13E3 is a single dominant source of the material for Sgr A* activity, and the
number density of its wind estimated near Sgr A* is roughly consistent with the X-ray data
limits of Baganoff et al. (2003)
\item the estimated mean angular momentum density $\lambda$ of the 
inflowing is most likely
between 1.68 and 2.16
in $R_{Schw}c $ units
\item for such an angular momentum and realistic values of the Bernoulli constant 
no stationary shock solutions exist and the
material piles up close to the black hole, forming a (non-stationary) ring/torus
\item for $\lambda = 2.16$ the inflow stops at $\sim 10 R_{Schw}$ and the 
radiation
spectrum (including bremsstrahlung, synchrotron and Comptonized radiation) 
emitted by the material above this radius is roughly consistent with the 
persistent emission from Sgr A* (Except of the radio part of the spectrum.)
\item for $\lambda = 1.68$ inflow stops at $\sim 4 R_{Schw}$, and the 
predicted persistent emission is too high for Sgr A*.
\item  the accretion rate predicted by our model is consistent with the
limits given by the measurements of the Faraday rotation if
magnetic field is highly tangled and makes many reversals, reducing the
Faraday Rotation measure by 200 times.

\end{itemize}

\section*{Acknowledgements}  
 
This work was  
supported in part by grants 2P03D~003~22 and 1P03D~008~29
of the Polish State  
Committee for Scientific Research (KBN). We gratefully acknowledge usefull 
discussions with Marek Abramowicz, Vasily Beskin, Mark Morris, and Marek Sikora.

\end{document}

\\  
This paper has been processed by the authors using the Blackwell  
Scientific Publications \LaTeX\  style file.  

\vfill\eject

\end{document}